\documentclass[useAMS,usenatbib]{mn2e}
\usepackage{epsf,graphicx,rotating,url}

\title[The black hole mass of 1H~0323$+$342]{On the black hole mass of the $\gamma$-ray emitting narrow-line Seyfert 1 galaxy 1H~0323$+$342}

\author[H. Landt et al.]{H. Landt$^1$\thanks{E-mail:hermine.landt@durham.ac.uk}, M. J. Ward$^1$, M. Balokovi\'c$^2$, D. Kynoch$^1$, T. Storchi-Bergmann$^3$, 
\newauthor C. Boisson$^4$, C. Done$^{1,5}$, J. Schimoia$^3$ and D. Stern$^6$ \\
$^1$Centre for Extragalactic Astronomy, Department of Physics, Durham University, South Road, Durham, DH1 3LE, UK \\ 
$^2$Cahill Center for Astronomy and Astrophysics, California Institute of Technology, 1216 E California Blvd, Pasadena, CA 91125, USA \\
$^3$Departamento de Astronomia, Universidade Federal do Rio Grande do Sul, IF, CP 15051, 91501-970 Porto Alegre, RS, Brazil \\
$^4$LUTH, Observatoire de Paris, CNRS, Universit\'e Paris Diderot, PSL Research University Paris, 5 place Jules Janssen, 92195 Meudon, France \\
$^5$ISAS, Japan Aerospace Exploration Agency, 3-1-1 Yoshinodai, chuo-ku, Sagamihara, Kanagawa 252-5210, Japan \\
$^6$Jet Propulsion Laboratory, California Institute of Technology, 4800 Oak Grove Drive, Pasadena, CA 91109, USA}

\begin{document}

\def\la{\mathrel{\hbox{\rlap{\hbox{\lower4pt\hbox{$\sim$}}}\hbox{$<$}}}}
\def\ga{\mathrel{\hbox{\rlap{\hbox{\lower4pt\hbox{$\sim$}}}\hbox{$>$}}}}

%emission line definitions
\font\sevenrm=cmr7
\def\SII{[S~{\sevenrm II}]}
\def\SIII{[S~{\sevenrm III}]}
\def\SVIII{[S~{\sevenrm VIII}]}
\def\SIX{[S~{\sevenrm IX}]}
\def\SXI{[S~{\sevenrm XI}]}
\def\SXII{[S~{\sevenrm XII}]}

\def\SiII{[Si~{\sevenrm II}]}
\def\SiIII{[Si~{\sevenrm III}]}
\def\SiVI{[Si~{\sevenrm VI}]}
\def\SiVII{[Si~{\sevenrm VII}]}
\def\SiVIII{[Si~{\sevenrm VIII}]}
\def\SiIX{[Si~{\sevenrm IX}]}
\def\SiX{[Si~{\sevenrm X}]}
\def\SiXI{[Si~{\sevenrm XI}]}

\def\FeII{Fe~{\sevenrm II}}
\def\FeIIf{[Fe~{\sevenrm II}]}

\def\OIII{[O~{\sevenrm III}]}

\def\HeII{He~{\sevenrm II}}

\def\cloudy{{\sevenrm CLOUDY}}
\def\xspec{{\sevenrm XSPEC}}
\def\heasoft{{\sevenrm HEASOFT}}
\def\donemodel{{\sevenrm OPTXAGNF}}
\def\newdonemodel{{\sevenrm OPTXCONV}}
\def\xrtpipeline{{\sevenrm XRTPIPELINE}}
\def\nupipeline{{\sevenrm NUPIPELINE}}
\def\nuproducts{{\sevenrm NUPRODUCTS}}
\def\nustardas{{\sevenrm NUSTARDAS}}

\date{Accepted ~~. Received ~~; in original form ~~}

\pagerange{\pageref{firstpage}--\pageref{lastpage}} \pubyear{2016}

\maketitle

\label{firstpage}

\begin{abstract}
Narrow-line Seyfert 1 galaxies have been identified by the {\it Fermi
  Gamma-Ray Space Telescope} as a rare class of $\gamma$-ray emitting
active galactic nuclei (AGN). The lowest-redshift candidate among them
is the source 1H~0323$+$342. Here we present quasi-simultaneous {\it
  Gemini} near-infrared and {\it Keck} optical spectroscopy for it,
from which we derive a black hole mass based on both the broad Balmer
and Paschen emission lines. We supplement these observations with a
{\it NuSTAR} X-ray spectrum taken about two years earlier, from which
we constrain the black hole mass based on the short timescale spectral
variability. Our multiwavelength observations suggest a black hole
mass of $\sim$2$\times$10$^7$~$M_\odot$, which agrees well with
previous estimates. We build the spectral energy distribution and show
that it is dominated by the thermal and reprocessed emission from the
accretion disc rather than the non-thermal jet component. A detailed
spectral fitting with the energy-conserving accretion disc model of
Done et al. constrains the Eddington ratio to $L/L_{\rm Edd} \sim 0.5$
for a (non-rotating) Schwarzschild black hole and to $L/L_{\rm Edd}
\sim 1$ for a Kerr black hole with dimensionless spin of $a^{\star}=
0.8$. Higher spin values and so higher Eddington ratios are excluded,
since they would strongly overpredict the observed soft X-ray flux.

\end{abstract}


\section{Introduction}

The majority of $\gamma$-ray emitting active galactic nuclei (AGN)
discovered by the {\it Fermi Gamma-Ray Space Telescope} and listed in
its third Large Area Telescope (LAT) catalogue \citep{Fermi3} are
blazars, evenly distributed between flat-spectrum radio quasars and BL
Lacertae objects. However, a very small number of $\gamma$-ray
emitting AGN are optically classified as narrow-line Seyfert 1s,
i.e. they have much lower optical luminosities than quasars and their
broad emission lines are relatively narrow with full widths at half
maximum (FWHM)$\la 2000$~km~s$^{-1}$. They usually also have very
strong emission lines from permitted \FeII~transitions in their
optical spectra \citep{Bor92}. Since the first discovery of
$\gamma$-ray emitting narrow-line Seyfert 1s \citep{Abdo09}, only
eight sources are known to date \citep{Fosch16a}. All of these sources
are radio-loud and their $\gamma$-ray emission is thought to be
produced via the external Compton (EC) mechanism whereby the
relativistic jet electrons upscatter a photon field external to the
jet, e.g. from the accretion disc, broad emission line region (BLR) or
dusty torus, to higher energies. This interpretation is also often
used to explain the $\gamma$-ray emission detected from broad-line
quasars.

The discovery of narrow-line Seyfert 1s as a class of $\gamma$-ray
emitting AGN is intriguing, since they generally reside in spiral
galaxies rather than in bright ellipticals which are usually the hosts
of radio-loud AGN with powerful relativistic jets. Furthermore, as a
class, the narrow-line Seyfert 1s tend to have lower black hole masses
and higher accretion rates relative to their Eddington limit compared
with the typical Seyfert 1 AGN. This means that the thermal accretion
disc spectrum and its Comptonised components are expected to dominate
over the jet emission at optical/UV wavelengths and X-ray
energies. This dominance is rarely seen over this entire frequency
range in the other $\gamma$-ray emitting blazar classes and so these
sources offer us the unique opportunity to study the connection
between jet and accretion power.

Among the $\gamma$-ray detected narrow-line Seyfert 1s, the source
1H~0323$+$342 is of particular interest, since it has the lowest
redshift \citep[$z=0.0629$;][]{Zhou07}. This not only means that its
host galaxy can be resolved by ground-based imaging \citep{Anton08,
  Leon14} and that due to its relatively high flux good
signal-to-noise (S/N) ratio observations can be obtained in relatively
short exposure times, but also that its black hole mass can be
reliably estimated from single-epoch spectra using several broad
emission lines. Its optical spectrum covers simultaneously the two
strongest Balmer lines, H$\alpha$ and H$\beta$, both for which
reliable black hole mass scaling relations exist
\citep[e.g.][]{Greene05, Xiao11, Bentz09, Bentz13} and a
cross-dispersed near-infrared (near-IR) spectrum with its large
wavelength coverage gives simultaneous observations of the two
strongest Paschen lines, Pa$\alpha$ and Pa$\beta$, for which a black
hole mass scaling relation has recently been presented by \citet{L11b,
  L13}. The black hole mass is a key ingredient for modelling the
accretion disc spectrum which in turn determines the accretion power
relative to the Eddington limit and the bolometric luminosity.

Here we present recent quasi-simultaneous optical and near-IR
spectroscopy of high quality (high S/N and moderate spectral
resolution), from which we derive a black hole mass based on both the
broad Balmer and Paschen emission lines. We supplement these
observations with a {\it NuSTAR} X-ray spectrum taken about two years
earlier, from which we constrain the black hole mass based on the
short timescale spectral variability.  This paper is organised as
follows. In Section 2, we describe the near-infrared, optical and
X-ray observations based on which we estimate the black hole mass as
detailed in Section 3. In Section 4, we construct the multiwavelength
spectral energy distribution (SED), which we fit with the
energy-conserving accretion disc model of \citet{Done12, Done13} in
order to constrain the Eddington ratio. Finally, in Section 5, we
summarise our main results and present our conclusions. Throughout
this paper we have assumed cosmological parameters $H_0 = 70$ km
s$^{-1}$ Mpc$^{-1}$, $\Omega_{\rm M}=0.3$, and
$\Omega_{\Lambda}=0.7$. Photon spectral indices have been defined as
$N(E) \propto E^{-\Gamma}$.

\section{The observations}

\subsection{The near-infrared spectroscopy} \label{nearir}

We observed the source 1H~0323$+$342 in queue mode with the Gemini
Near-Infrared Spectrograph \citep[GNIRS;][]{gnirs} at the Gemini North
8~m observatory in semester 2015B (Program ID: GN-2015B-FT-4) in the
framework of the recently initiated Fast Turnaround Program. The
observations were taken on September 16, 2015. There were no clouds
and the seeing was excellent. We used the cross-dispersed mode with
the short camera at the lowest resolution (31.7~l~mm$^{-1}$ grating),
thus covering the entire wavelength range of $0.9-2.5$~$\mu$m without
inter-order contamination. We chose a slit of $0.45\times7''$. This
set-up gives an average spectral resolution of full width at
half-maximum (FWHM) $\sim 265$~km~s$^{-1}$. The on-source exposure
time was $6\times90$~sec at an average airmass of $\sec z=1.037$,
which resulted in an average continuum $S/N \sim 40$, 70 and 90 in the
$J$, $H$ and $K$~bands, respectively. Since the source is too extended
for the relatively small slit length, we nodded off onto a blank patch
of sky for the background subtraction.

Before the science target, we observed the nearby (in position and air
mass) A2~V star HIP~16168 that has accurate near-IR magnitudes. We
used this standard star to correct our science spectrum for telluric
absorption and for flux calibration. Flats and arcs were taken after
the science target.The data were reduced using the Gemini/IRAF package
(version 1.13) with GNIRS specific tools \citep{gnirssoft}. The data
reduction steps included preparation of calibration and science
frames, processing and extraction of spectra from science frames,
wavelength calibration of spectra, telluric correction and
flux-calibration of spectra, and merging of the different orders into
a single, continuous spectrum. The spectral extraction width was
adjusted interactively for the telluric standard star and the science
source to include all the flux in the spectral trace. The final
spectrum was corrected for Galactic extinction using the IRAF task
\mbox{\sl onedspec.deredden} with an input value of $A_{\rm V}=0.706$,
which we derived from the Galactic hydrogen column densities published
by \citet{DL90}. The result is shown in Fig. \ref{gnirsspec}.

\subsection{The optical spectroscopy} \label{optical}

We obtained an optical spectrum of the source 1H~0323$+$342 on
February 14, 2016, with the Low Resolution Imaging Spectrometer
\citep[LRIS;][]{lris} mounted on the Keck 10~m telescope. The weather
was photometric with a seeing of $\sim 0.6$~arcsec. We used the
600/4000 and 400/8500 gratings for the blue and red arms,
respectively, with the $1''$~slit. This set-up gives a relatively
large spectral coverage of $\sim 3100 - 10300$~\AA, with a very small
spectral gap of $\sim 40$~\AA~between the two arms. The average
spectral resolution is FWHM $\sim 300$~km~s$^{-1}$, similar to that of
our near-IR spectroscopy. The slit was rotated to the parallactic
angle, but note that the LRIS has an atmospheric dispersion
corrector. The on-source exposure time was $300$~sec at an average
airmass of $\sec z=1.65$, which resulted in an average continuum $S/N
\sim 60$. The data were reduced using standard longslit routines from
the IRAF software package. The extracted spectrum was flux-calibrated
using the standard stars G191B2B and HZ~44 with fluxes as given in
\citet{Massey90}. The final spectrum was corrected for Galactic
extinction as was done for the near-IR spectrum (see Section
\ref{nearir}). The result is shown in Fig. \ref{lrisspec}. We note
that based on the observed wavelength of the forbidden narrow emission
lines and narrow components of the broad emission lines in both our
near-IR and optical spectra, we get a redshift of $z=0.0625$, which
differs by $\sim 120$~km~s$^{-1}$ from the value of $z=0.0629$
published by \citet{Zhou07}. At the spectral resolution of our data
this difference is significant, since the wavelength position of the
emission line peak can be determined with sub-pixel accuracy.

The optical and near-IR spectrum have a considerable wavelength region
of overlap at their respective red and blue ends, which we can use to
test if the source flux has varied between the two observing
epochs. Importantly, this overlap wavelength region covers the strong
forbidden narrow emission line \SIII~$\lambda 9531$. For this line we
measure an integrated flux of 1.25$\times$10$^{-15}$ and
8.92$\times$10$^{-16}$~erg~s$^{-1}$~cm$^{-2}$ for the near-IR and
optical spectrum, respectively. The difference in continuum flux in
the overlap wavelength region is similar ($\sim 40\%$), with the
optical spectrum having again a lower flux than the near-IR
spectrum. Therefore, flux calibration issues rather than genuine
source variability are favoured as the cause for the flux misalignment
between the two spectra. In order to further check the absolute flux
calibration of the optical spectrum, we have compared the flux of the
strong forbidden narrow emission line \OIII~$\lambda 5007$ to that
observed in the optical spectrum published by \citet{Marcha96}. Their
spectrum was obtained in November 1992 with the Multiple Mirror
Telescope (MMT) 4.5 m on Mt. Hopkins, Arizona, USA, using a
$1.5''$~slit oriented at parallactic angle and the 150 l~mm$^{-1}$
grating. This set-up resulted in a spectral resolution of $\sim
1200$~km~s$^{-1}$ over the wavelength range of $\sim 3770 -
8683$~\AA. After correcting the MMT spectrum for Galactic reddening
and subtracting the \FeII~emission in both spectra as described in
Section \ref{nearirbh}, we measure an integrated flux of
1.06$\times$10$^{-14}$ and
1.50$\times$10$^{-14}$~erg~s$^{-1}$~cm$^{-2}$ for the \OIII~$\lambda
5007$ line in the Keck and MMT optical spectrum, respectively. The
difference between the two is $\sim 40\%$, similar to what we found
when comparing the Keck optical spectrum with our near-IR
spectrum. Therefore, for the following analysis, we have scaled the
optical spectrum up to the flux level of the near-IR spectrum. We note
that \citet{Berton16} have also recently obtained an optical
spectrum. The \OIII~$\lambda 5007$ line luminosity that they measure
in their 2014/15 spectrum from the Asiago 1.2~m telescope is $\sim
10\%$ lower than our measurement in the Keck spectrum.

\subsection{The X-ray spectroscopy}

\subsubsection{NuSTAR} \label{nustar}

The source 1H~0323$+$342 was observed with the {\it Nuclear
  Spectroscopic Telescope Array} \citep[{\it NuSTAR};][]{nustar}
between March 15-18, 2014 for a total duration of 198.72~ks. We
processed the data using the \nupipeline~script available in
\nustardas~version 1.4.1 (\heasoft~version 1.16) with the calibration
database CALDB version 20150316. Occasional high count rates were
filtered out using the {\tt SAA=Strict} filter, reducing the total
exposure time by $\la10\%$. The sum of good-time intervals after event
cleaning and filtering is 91~ks for both focal plane modules A (FPMA)
and B (FPMB). For each module, we extracted source spectra from
circular regions with $60''$ radius centered on the peak of the
emission, while the background spectra were extracted from source-free
regions of the same detector, toward the edge of the field of
view. Background accounts for $\la6\%$ of the counts in the source
region within the 3--79\,keV bandpass. We obtain a $S/N \sim 100$ and
$\sim 70$ per module for the 3--10~keV and 10--79~keV bands,
respectively.

Source spectra, light curves and response files were generated using
the \nuproducts~script. The light curves for the energy bands
3--10~keV and 10--79~keV are shown in Fig. \ref{nustarlc}, top
panel. We used a time binning of 30-min. in order to have at least two
time bins per {\it NuSTAR} orbit (of $\simeq90$ min.). We omitted bins
containing less than 3 min. of source exposure. The point spread
function (PSF)-corrected total source count rate was fluctuating
around 0.33~cts/s, except for a short period near the middle of the
observation when it flared up to $\simeq0.5$~cts/s. In
Fig. \ref{nustarlc}, bottom panel, we show the hardness ratio as a
function of time to demonstrate that the spectrum did not
significantly change during this brief period of increased flux. For
the X-ray spectral analysis we describe in the following we used the
time-averaged spectra binned to a minimum of 50 counts per energy
bin. For the analysis of the SED in Section \ref{sed}, we co-added the
FPMA and FPMB spectra using standard \heasoft~{\tt ftools} and used 12
wide energy bins over the total 3--79~keV band to aid visibility in
Fig. \ref{sedplot}.

We first analyzed the two {\it NuSTAR} spectra separately using
\xspec~\citep{xspec}. We fitted the FPMA and FPMB spectra
simultaneously, without co-adding. We assumed the following three
models: a single power-law, a log-parabolic model ($F(E) \propto
E^{-\alpha-\beta \log E}$) and the sum of two power-laws. All models
assumed a fixed redshift for the source of $z=0.0625$, a fixed
Galactic hydrogen column density of $N_{\rm H}=14.62 \times
10^{20}$~cm$^{-2}$ \citep{DL90} and a cross-normalization factor,
which was allowed to vary in the spectral fits. We selected the
best-fit model by requiring an $F$-test probability $>99.99\%$ that
the $\chi^2$ value of the model with the larger number of free
parameters represents an improvement. We found that a single power-law
model fits the data best, with $\chi^2=575.1$ for 515 degrees of
freedom (d.o.f.) giving a reduced $\chi^2$ value of
$\chi^2_{\nu}=1.12$. The best-fit photon index for this model is
$\Gamma=1.80\pm0.01$, where the uncertainty is given as the $1\sigma$
confidence interval. The resulting 2--10~keV flux is
8.10$\times$10$^{-12}$~erg~s$^{-1}$~cm$^{-2}$. Replacing the power-law
continuum with the more flexible log-parabolic model or the sum of two
power-laws does not give a statistically significant improvement. We
obtain a $\chi^2=568.6$ for 514 d.o.f. and a $\chi^2=568.6$ for 513
d.o.f. for the former and latter model, respectively, resulting in
$F$-test probabilities of $98.4\%$ and $94.6\%$,
respectively. However, it is worth mentioning that the fit with two
power-laws gives spectral indices of $\Gamma_1 \sim 2$ and $\Gamma_2
\sim 1.2-1.6$ below and above a break energy of $E_{\rm break} \sim
25$~keV, respectively, which correspond well to the typical photon
indices for coronal and jet contributions observed in the X-ray
spectra of Seyferts, in particular of narrow-line Seyfert 1s (see also
Section \ref{jet}).

We have tested the data for the presence of other spectral features
commonly observed in the hard X-ray spectra of AGN, namely, the
Compton hump and the iron line at 6.4~keV, by adding two components to
the power-law continuum. We used the {\tt pexrav} model \citep{pexrav}
to represent the reprocessed continuum with most parameters fixed
(high-energy cutoff at 1~MeV, inclination at $60^\circ$, elemental
abundances at Solar values), and a narrow, unresolved Gaussian to
represent the emission line. We find that the contributions of these
two components are small and that this model does not constitute a
statistically significant improvement in comparison to the single
power-law ($\chi^2=567.6$ for 513 d.o.f., resulting in an $F$-test
probability of $96.6\%$). Finally, we note that for all models we find
that the cross-normalization factor (FPMB/FPMA) is $1.06\pm0.02$,
which is on the high side but still within expectations from {\it
  NuSTAR} calibration \citep{Madsen15}.

\subsubsection{Swift} \label{swift}

\begin{table*}
\caption{\label{swiftlog} {\it Swift} XRT and UVOT observations}
\begin{tabular}{lcccccc}
\hline
Observation & ObsID & Exposure & Source & $\Gamma$ & $f_{(2-10{\rm keV})}$   & $\chi^2_{\nu}$/dof \\ 
Date        &       & (s)      & Counts &          & (erg/s/cm$^2$) & \\
(1)         & (2)   & (3)      & (4)    & (5)      & (6)            & (7) \\
\hline
2013 Sep 20 & 00036533044 & 3798 & 1222 & 1.96$\pm$0.05 & 8.35e$-$12 & 0.80/52 \\
2014 Dec 10 & 00036533052 & 2972 & 1365 & 1.94$\pm$0.05 & 1.10e$-$11 & 0.95/59 \\
2015 Sep 17 & 00036533064 & 1636 &  670 & 1.99$\pm$0.08 & 9.79e$-$12 & 0.98/30 \\
\hline
Observation & V        & B        & U        & UVW1     & UVM2     & UVW2 \\
Date        & (mag)    & (mag)    & (mag)    & (mag)    & (mag)    & (mag) \\
(1)         & (8)      & (9)      & (10)     & (11)     & (12)     & (13) \\
\hline
2013 Sep 20 & 15.62$\pm$0.06 & 16.10$\pm$0.04 & 15.24$\pm$0.04 & 15.36$\pm$0.05 & 15.71$\pm$0.07 & 15.55$\pm$0.05 \\
2014 Dec 10 & 15.51$\pm$0.05 & 16.02$\pm$0.04 & 15.09$\pm$0.04 & 15.38$\pm$0.05 & 15.74$\pm$0.06 & 15.70$\pm$0.05 \\
2015 Sep 17 & 15.70$\pm$0.09 & 16.25$\pm$0.06 & 15.35$\pm$0.06 & 15.57$\pm$0.08 & 15.78$\pm$0.09 & 15.78$\pm$0.07 \\
\hline
\end{tabular}

\parbox[]{13.2cm}{The columns are: (1) date of observation; (2)
  observation ID; for the XRT X-ray observations (3) filtered live
  exposure time; (4) extracted source counts; (5) photon index; (6)
  observed flux in the range $2-10$ keV; and (7) reduced $\chi^2$ and
  number of degrees of freedom for a single power-law fit with a fixed
  Galactic hydrogen column density of $N_{\rm H}=14.62 \times
  10^{20}$~cm$^{-2}$; for the simultaneous UVOT observations the
  observed (absorbed) Vega magnitudes in the (8) V filter
  ($\lambda_{\rm eff}=5402$~\AA), (9) B filter ($\lambda_{\rm
    eff}=4329$~\AA), (10) U filter ($\lambda_{\rm eff}=3501$~\AA),
  (11) UVW1 filter ($\lambda_{\rm eff}=2634$~\AA), (12) UVM2 filter
  ($\lambda_{\rm eff}=2231$~\AA), and (13) UVW2 filter ($\lambda_{\rm
    eff}=2030$~\AA). We quote all errors at the 1$\sigma$ level.}

\end{table*}

Since our near-IR/optical spectroscopy and the {\it NuSTAR} X-ray
spectrum are separated in time by about two years, we have used
archival {\it Swift} observations, which have simultaneous optical/UV
magnitudes and X-ray spectra, to check for extreme flux variability
between the two epochs. {\it Swift} observed the source 1H~0323$+$342
on September 17, 2015, i.e. only one day after the near-IR
spectroscopy was taken, but there are no observations very close in
time with the \mbox{\it NuSTAR} spectroscopy. Therefore, we have used
those two {\it Swift} observations that are the closest in time with
it, namely, the data taken on September 20, 2013, and on December 10,
2014, i.e. about six months earlier and about nine months later,
respectively. Within the {\it Swift} archive, we used data collected
with the X-ray Telescope (XRT) in photon counting mode. We reprocessed
the initial event files with the \xrtpipeline~(version 0.13.2) using
standard settings and the latest known calibration files. Source
spectra were extracted from circular regions corresponding to an
encircled energy of $\sim 90\%$ at 1.5 keV. Background spectra were
taken from a circular region with a radius roughly three times as
large as that of the source and offset from the source position. The
background-subtracted spectra were fit using \xspec, with the response
matrices from the calibration database. The source spectra were binned
to a minimum of 20 counts per energy bin in order to apply the
$\chi^2$ minimization technique. We fit the data with two different
models, namely, a single power-law and a broken power-law. The
hydrogen column density was fixed to the Galactic value and in the
case of the single power-law fits also allowed to vary. The best-fit
model was in all cases a single power-law with a fixed Galactic
hydrogen column density. The relevant parameter values for the X-ray
fits together with the simultaneous optical/UV magnitudes from the
Ultraviolet Optical Telescope (UVOT) on-board {\it Swift} are reported
in Table \ref{swiftlog}.

We find that the {\it Swift} X-ray flux in the 2-10~keV energy range
changed by $\sim 30\%$ between the two observing epochs before and
after the \mbox{\sl NuSTAR} X-ray spectroscopy, with the earlier epoch
having a very similar X-ray flux to the \mbox{\sl NuSTAR} spectrum and
the later one a value only $\sim 10\%$ higher than the observing epoch
close in time with the near-IR spectroscopy. Furthermore, the average
between the two is very similar to the X-ray flux of the observing
epoch corresponding to the near-IR spectrum. Neither is strong flux
variability observed in the optical/UV. Significant flux changes are
detected in the $B$, $U$, $UVW1$ and $UVW2$ filters, but only by $\sim
20-30\%$ at the $2-3\sigma$~level. Therefore, we have co-added the
three {\it Swift} X-ray spectra and performed again the spectral
fits. This time, the best-fit model was a broken power-law with a
fixed Galactic hydrogen column density ($\chi^2_{\nu}=0.87$ for 124
d.o.f.), showing clearly the soft X-ray excess typical of narrow-line
Seyfert 1s in addition to the hard power-law. The resultant spectral
indices in the soft and hard X-ray bands are $\Gamma_{\rm
  soft}=2.22^{+0.06}_{-0.05}$ and $\Gamma_{\rm
  hard}=1.72^{+0.10}_{-0.12}$, respectively, for a break energy of
$E_{\rm break}=2.04^{+0.41}_{-0.31}$~keV. The $0.3-10$~keV flux is
$1.69 \times 10^{-11}$~erg~s$^{-1}$~cm$^{-2}$, which points at the
source being in a high state and intermediate between the first and
second flare investigated by \citet{Paliya14} in their 2008-2013 {\it
  Swift} XRT light curve (see their Table 5).

\section{Estimates of the black hole mass}

In this section, we estimate the black hole mass in the source
1H~0323$+$342 using two different quantities that scale with it,
namely, the virial product between the width of a broad emission line
and the continuum luminosity, which serves as a proxy for the BLR
radius, and the short timescale variability in the X-ray band. The
first method assumes that the dynamics of the gas in the BLR is
dominated by gravitational forces and uses the virial theorem to
calculate the black hole mass as:

\begin{equation}
\label{virial}
M_{\rm BH} = f \frac{R \Delta V^2}{G},
\end{equation}

\noindent
where $R$ is the radial distance of the BLR gas from the black hole,
$\Delta V$ is the velocity dispersion of the gas, $G$ is the
gravitational constant and $f$ is a scaling factor that depends on the
(unknown) dynamics and geometry of the BLR. The BLR radius can be
directly measured through reverberation mapping, a technique which
determines the light-travel time-delayed lag with which the flux of
the BLR responds to changes in the {\it ionising} continuum
flux. However, since reverberation mapping campaigns are observing
time intensive, the so-called radius-luminosity relationship is used
to estimate the BLR sizes for large samples of AGN from single-epoch
spectra. As has been shown \citep[e.g.][]{Pet93, Wan99, Kaspi00,
  Bentz09, L11b}, the BLR lags obtained from reverberation mapping
campaigns correlate with the optical, UV and near-IR luminosity (of
the ionising component) largely as expected from simple
photoionisation arguments.

The second method assumes that the X-ray variability properties, such
as time scales and amplitude, of all sources (galactic and
extragalactic) that host a black hole are determined by its mass; the
larger the black hole mass, the larger the size of the X-ray emitting
region and so the longer the timescales on which the X-ray emission
varies, leading to smaller variability amplitudes \citep{Barr86,
  Green93, Nandra97, McHardy06}.

\begin{table*}
\caption{\label{bhmass} Estimates of the black hole mass using different methods}
\begin{tabular}{llcl}
\hline
Wavelength & Measurements & $M_{\rm BH}$        & Reference \\
           &              & (10$^7$ M$_\odot$)  & \\
\hline
Near-IR   & $\log \nu L_{\rm 1\mu m}=43.92$~erg~s$^{-1}$  & 2.0$^{+0.8}_{-0.6}$           & eq. (2) of \citet{L13} \\ 
          & FWHM(Pa$\alpha$)=1120~km~s$^{-1}$             & & \\ 
\hline
Near-IR   & $\log \nu L_{\rm 1\mu m}=43.92$~erg~s$^{-1}$  & 1.8$^{+1.3}_{-0.7}$           & eq. (3) of \citet{L13} \\ 
          & $\sigma$(Pa$\alpha$)=875~km~s$^{-1}$          & & \\ 
\hline
Near-IR   & $\log L_{\rm Pa\alpha}=41.46$~erg~s$^{-1}$    & 1.0$\pm$0.2                   & eq. (9) of \citet{Kim10} \\
          & FWHM(Pa$\alpha$)=1120~km~s$^{-1}$             & & \\
\hline
Optical   & $\log \nu L_{\rm 5100\AA}=44.05$~erg~s$^{-1}$ & 1.5$^{+0.7}_{-0.5}$$^\star$   & Table 7 of \citet{Mejia16} \\
          & FWHM(H$\alpha$)=1412~km~s$^{-1}$              & & \\
\hline
Optical   & $\log L_{\rm H\alpha}=42.44$~erg~s$^{-1}$     & 0.6$^{+0.4}_{-0.2}$$^\star$   & Table 7 of \citet{Mejia16} \\
          & FWHM(H$\alpha$)=1412~km~s$^{-1}$              & & \\
\hline
Optical   & $\log \nu L_{\rm 5100\AA}=44.05$~erg~s$^{-1}$ & 2.2$^{+0.8}_{-0.6}$$^\star$   & Table 14 of \citet{Bentz13} \\
          & FWHM(H$\beta$)=1437~km~s$^{-1}$               & & \\
\hline
Optical   & $\log L_{\rm H\beta}=42.02$~erg~s$^{-1}$      & 1.2$^{+0.8}_{-0.5}$$^\star$   & Table 2 of \citet{Greene10} \\
          & FWHM(H$\beta$)=1437~km~s$^{-1}$               & & \\ 
\hline
X-ray/    & $\log L_{\rm 2-10keV}=43.97$~erg~s$^{-1}$     & 2.2$^{+1.8}_{-1.0}$$^\star$   & Table 2 of \citet{Greene10} \\
Optical   & FWHM(H$\beta$)=1437~km~s$^{-1}$               & & \\
\hline
X-ray     & $\sigma_{\rm rms}^2=0.007\pm0.005$ (20~ks)    & 1.0$^{+3.3}_{-0.5}$$^\dagger$ & Table 3 of \citet{Ponti12} \\
\hline
X-ray     & $\sigma_{\rm rms}^2=0.005\pm0.003$ (40~ks)    & 1.7$^{+3.3}_{-0.8}$$^\dagger$ & Table 3 of \citet{Ponti12} \\
\hline
\end{tabular}

\parbox[]{13.7cm}{$^\star$ The $1\sigma$ error is derived from the
  instrinsic scatter rather than the errors on the best-fit parameter
  values.\\ 
$^\dagger$ The $1\sigma$ error includes the measurement errors in
  addition to the errors on the best-fit parameter values.}

\end{table*}

\subsection{Near-infrared and optical spectroscopy} \label{nearirbh}

\begin{figure*}
\centerline{
\includegraphics[clip=true, bb=20 400 570 710, scale=0.9]{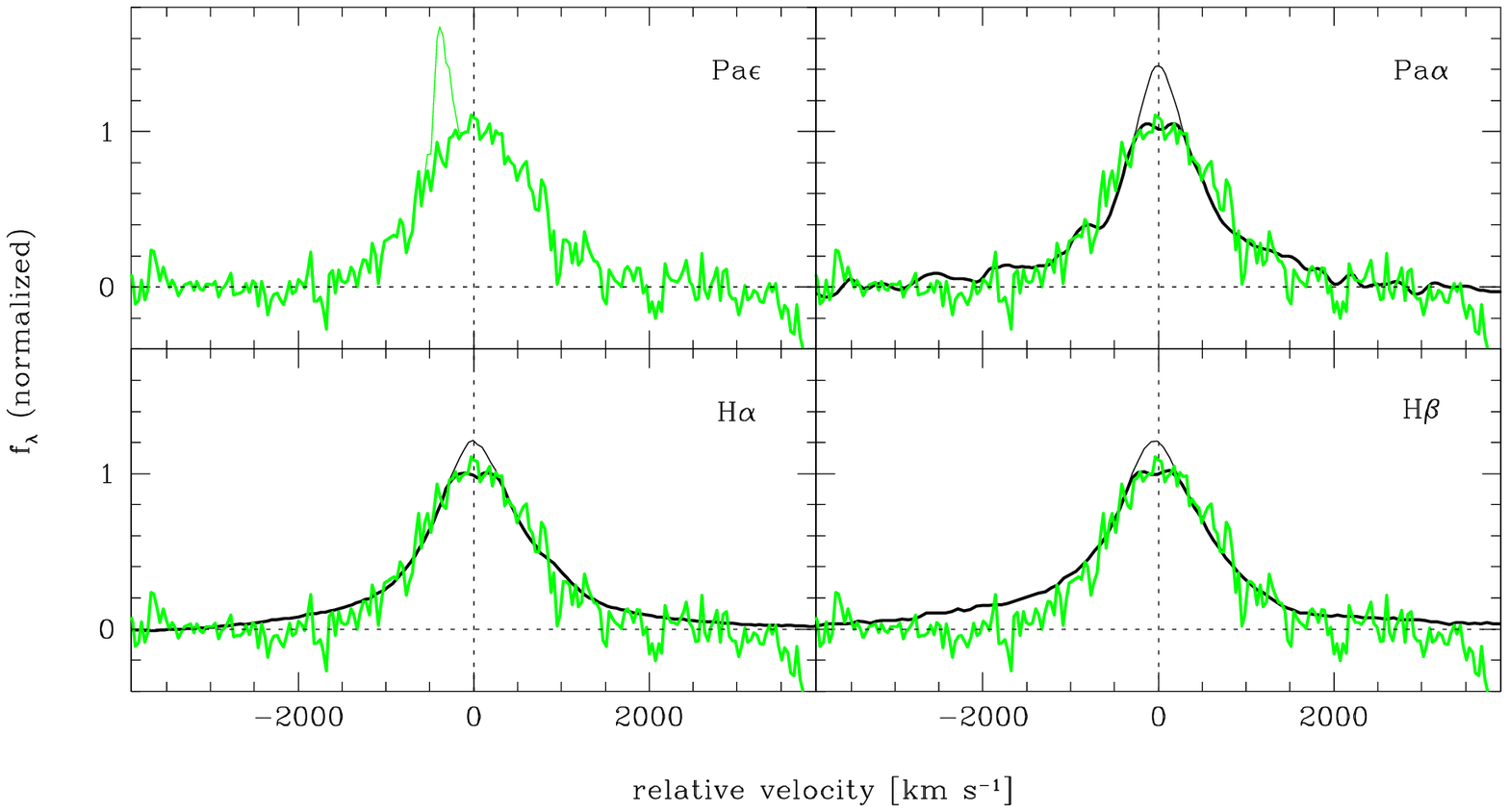}
}
\caption{\label{PaAPaE} The profile of the Pa$\epsilon$ emission line
  (thick green lines in upper left-hand panel; here taken from the
  Keck LRIS optical spectrum) is blended with the forbidden narrow
  emission line \SIII~$\lambda 9531$ (thin green lines) but its narrow
  component is absent. After removing the largest possible flux
  contribution from the narrow emission line region (thin black
  lines), the resulting profile of the broad component of the
  Pa$\alpha$, H$\alpha$ and H$\beta$ emission lines (thick black lines
  in upper right-hand, lower left-hand and lower right-hand panels,
  respectively) is similar to that of the Pa$\epsilon$ emission line
  (thick green lines).}
\end{figure*}

We first estimate the black hole mass using the near-IR relationship
presented by \citet{L13}. This relationship is based on the virial
product between the 1~$\mu$m continuum luminosity and the width (FWHM
or line dispersion) of the strongest Paschen broad emission lines,
Pa$\alpha$ or Pa$\beta$. As detailed by these authors, the main
advantage of the near-IR virial product over the optical one is the
reliable measurement of its quantities; both Pa$\alpha$ and Pa$\beta$
are observed to be unblended and the continuum around 1~$\mu$m is free
of major contaminating components. Host galaxy starlight has its
emission maximum at $\sim 1$~$\mu$m, but its contribution is usually
negligible in luminous AGN if the near-IR spectrum was obtained
through a small slit. In our case, we can estimate the host galaxy
flux at 1$~\mu$m enclosed by our spectral aperture using the imaging
results of \citet{Leon14}. These authors give surface brightness
estimates in the $R$~band for both a bulge and a bulge plus disc model
of the host galaxy, which we use to scale the S0 galaxy template of
\citet{Pol07}. In this way, we estimate that the luminosity of the
host galaxy at 1$~\mu$m in our near-IR spectrum lies a factor of $\ga
70$ below the total luminosity.

The Gemini GNIRS near-IR spectrum covers simultaneously both the
Pa$\alpha$ and Pa$\beta$ emission lines as well as the rest-frame
1~$\mu$m wavelength region. However, the Pa$\beta$ emission line is
strongly affected by atmospheric absorption since it lies at the end
of the atmospheric window. Therefore, in the following, we estimate
the black hole mass using only the Pa$\alpha$ broad emission line, for
which we measure the line width as the FWHM, i.e. we use specifically
eq. (2) in \citet{L13}. The measurement of the 1~$\mu$m continuum
luminosity is straightforward and we get a value of $\log \nu L_{\rm
  1\mu m}=43.92$~erg~s$^{-1}$. However, in order to correctly measure
the width of the Pa$\alpha$ broad component, we need to first separate
it from the narrow component. Based on absent narrow components
observed for the higher-order Paschen emission lines, \citet{L14} have
suggested that the correct approach to this is to subtract the largest
possible flux contribution from the narrow emission line region. This
contribution is obtained by first estimating the FWHM of a strong
forbidden narrow emission line and then fitting a Gaussian of this
width to the top part of the total emission line profile. We have
applied this method here and show the result in Fig. \ref{PaAPaE}. In
the source 1H~0323$+$342, we find that already the Pa$\epsilon$
emission line, which is blended with the strong forbidden narrow
emission line \SIII~$\lambda 9531$, has no narrow component (thick
green line in Fig. \ref{PaAPaE}, upper left-hand panel). A Gaussian
fit to the \OIII~$\lambda 5007$ emission line observed in the optical
spectrum gives a FWHM=294~km~s$^{-1}$. Subtracting a Gaussian with
this width from the top part of the total Pa$\alpha$ emission line
(thin black lines) leaves a broad component with a similar profile to
that observed for the Pa$\epsilon$ emission line (Fig. \ref{PaAPaE},
upper right-hand panel). We measure a FWHM=1120~km~s$^{-1}$ for the
Pa$\alpha$ broad component, which results in a black hole mass of
(2.0$^{+0.8}_{-0.6}$)$\times$10$^7$~solar masses (see also Table
\ref{bhmass}).

We next estimate the black hole mass using the latest scaling
relations based on the optical virial product between the ionising
5100~\AA~continuum luminosity and the width of the strongest Balmer
broad emission lines H$\alpha$ and H$\beta$. All these three
quantities are covered simultaneously by our Keck LRIS optical
spectrum. The measurement of the 5100~\AA~continuum luminosity is
straightforward and we get a value of $\log \nu L_{\rm
  5100\AA}=44.05$~erg~s$^{-1}$ (in the scaled-up spectrum as described
in Section \ref{optical}). After subtracting the narrow component of
the H$\alpha$ emission line in a similar way as we did for the
Pa$\alpha$ emission line (Fig. \ref{PaAPaE}, lower left-hand panel),
we measure a FWHM=1412~km~s$^{-1}$ for the H$\alpha$ broad component.
Using the recently recalibrated black hole mass relationship of
\citet{Mejia16}, specifically the calibration for the local continuum
fit corrected for small systematic offsets (see their Table 7), we
estimate the black hole mass to be
(1.5$^{+0.7}_{-0.5}$)$\times$10$^7$~solar masses (see also Table
\ref{bhmass}).

As is typical for narrow-line Seyfert 1 galaxies, the emission from
permitted \FeII~transitions is very strong in the source
1H~0323$+$342. This emission needs to be modelled and subtracted in
order to reliably measure the width of the H$\beta$ emission line,
since the numerous optical \FeII~multiplets form a pseudo-continuum
around the line and blend in with its red wing. We used the template
based on the optical spectrum of I~Zw~1 published by \citet{Veron04}
and available in electronic format to model and subtract the
\FeII~emission. The method generally used to subtract the
\FeII~emission from optical spectra was first introduced by
\citet{Bor92}. It consists of creating a spectral sequence by
broadening (by convolution with Gaussians) and scaling of the
\FeII~template, which is subsequently packed together into a
three-dimensional cube. This cube is then subtracted from a cube
consisting in all three dimensions of the object's spectrum. But, as
noted by \citet{L08a} and \citet{Ves05}, it can be rather difficult to
decide by eye unambiguously which pair of width and strength of the
\FeII~template gives the cleanest subtraction, and so it is necessary
to constrain a priori the width of the \FeII~template. Following
\citet{L08a}, we have done this by using the width of the unblended
near-IR iron emission line \FeII~1.0502 $\mu$m. For this we measure a
value of FWHM=1034~km~s$^{-1}$, which is similar to the line width of
FWHM=1100~km~s$^{-1}$ used for the \FeII~template. Therefore, in this
case we did not need to broaden the \FeII~template but only to scale
it. In this way, we achieved a satisfactory \FeII~subtraction around
the H$\beta$ line. After subtracting its narrow component in a similar
way as we did for the H$\alpha$ and Pa$\alpha$ emission lines
(Fig. \ref{PaAPaE}, lower right-hand panel), we measure a
FWHM=1437~km~s$^{-1}$ for the H$\beta$ broad component. Using the
radius-luminosity relationship for the H$\beta$ line of
\citet{Bentz13} derived from optical reverberation mapping results,
specifically their calibration 'Clean2+ExtCorr' cleaned for bad time
lags and corrected for internal extinction (see their Table 14), and
assuming a geometrical scaling factor of $f=1.4$ appropriate for FWHM
measures \citep{Onken04}, we derive a black hole mass of
(2.2$^{+0.8}_{-0.6}$)$\times$10$^7$~solar masses (see also Table
\ref{bhmass}).

\subsection{X-ray variability} \label{xrayvar}

The short timescale X-ray variability is best quantified by deriving
the power density spectrum and determining its amplitude and break
frequency, i.e. the frequency at which the spectral slope changes. But
such an analysis is very difficult with unevenly sampled data as
afforded by low-Earth orbit satellites such as {\it
  NuSTAR}. Therefore, we have considered here instead the normalized
excess variance, $\sigma_{\rm rms}$, which was first introduced by
\citet{Nandra97} as an X-ray variability measure. This quantity has
been shown to correlate well with black hole mass, most recently by
\citet{Ponti12} who used data obtained with {\it XMM-Newton}. Because
the bandpasses of {\it XMM-Newton} and {\it NuSTAR} differ, count
rates of the former are dominated by low-energy photons. However,
considering only the $3-10$~keV band of {\it NuSTAR}, the differences
are minimized and we assume that the scaling relations hold for
variability statistics based on {\it NuSTAR} count rates as well. 

We have computed the $3-10$~keV light-curve (see Fig. \ref{nustarlc})
and calculated the normalized excess variance and its $1\sigma$ error
following \citet{Vaughan03} using segments of total length of 20~ks
and 40~ks. Averaging between FPMA and FPMB, we obtain values of
$\sigma_{\rm rms}^2=0.007\pm0.005$ and $0.005\pm0.003$ for the 20~ks
and 40~ks cases, respectively. Using the relationships between
$\sigma_{\rm rms}^2$ and black hole mass published by \citet{Ponti12}
for their reverberation-mapped AGN sample and listed in their Table 3,
we obtain a black hole mass of (1.0$^{+3.3}_{-0.5}$)$\times$10$^7$ and
(1.7$^{+3.3}_{-0.8}$)$\times$10$^7$~solar masses for the 20~ks and
40~ks cases, respectively (see also Table \ref{bhmass}). These values
are in good agreement with our results from the near-IR and optical
spectroscopy presented above. However, if we use instead the
relationships for their entire sample and listed in their Table 5, we
obtain considerably larger values, namely, a black hole mass of
2.1$\times$10$^7$ and 4.3$\times$10$^7$~solar masses for the 20~ks and
40~ks binning, respectively. In particular, the latter value is
inconsistent with our near-IR/optical spectroscopy results. However,
we note that our {\it NuSTAR} light curve has a time binning of
30~min. to minimise the impact of the orbit gaps and maximise S/N, but
this means that, compared to the 250~sec. binning used by
\citet{Ponti12}, we may underestimate the excess variance and so
overestimate the black hole mass.

\subsection{Discussion}

Our three estimates of the black hole mass in the source 1H~0323$+$342
based on the ionising continuum luminosity and the width of an
hydrogen broad emission line give a very small range of values of
$\sim 1.5 - 2.2 \times 10^7$~solar masses, with an average value of
$\sim 2 \times 10^7$~solar masses. This excellent agreement between
the three estimates is suprising, given that the relationships upon
which they are based have uncertainties of the order of $\sim 40-50\%$
at the $1\sigma$~level. The agreement between black hole mass
estimates from the broad emission lines and those from the short-term
X-ray variability, which lie in the range of $\sim 1.0 - 1.7 \times
10^7$~solar masses, are also in reasonable agreement with each other,
although we have used now a completely different method.

There are also other methods that can be used to estimate the black
hole mass, which are based on the dispersion instead of the FWHM of
the broad emission line \citep{L13}, and the line and X-ray luminosity
instead of the ionising continuum luminosity \citep{Kim10,
  Greene10}. We have also considered these alternative methods and
list the results in Table \ref{bhmass}. The relationships based on the
line dispersion and X-ray luminosity give results in good agreement
with our previous estimates. However, using the line luminosities of
Pa$\alpha$, H$\alpha$ and H$\beta$ we obtain black hole masses in the
range of $\sim 0.6-1.2 \times 10^7$~solar masses, which are a factor
of $\sim 2$ smaller than our previous estimates, but closer to those
obtained from the X-ray variability.

The black hole mass in the source 1H~0323$+$342 was previously
estimated by other authors. \citet{Zhou07} used the H$\beta$ line
luminosity and 5100~\AA~continuum luminosity together with the FWHM of
H$\beta$ and estimated the black hole mass in the range of $\sim 1-3
\times 10^7$~solar masses, which is consistent with our
results. \citet{Leon14} also used the H$\beta$ line luminosity and
5100~\AA~continuum luminosity together with the FWHM of H$\beta$ and
obtained values in the range of $\sim$0.8-2$\times 10^7$~solar
masses. In addition, they estimated the black hole mass based on the
luminosity of the host galaxy bulge and obtained values a factor of
$\sim 10$ higher ($\sim$3-5$\times 10^8$~solar masses). This is in
line with the well-known discrepancy between the black hole mass
estimates for narrow-line Seyfert 1s based on their broad emission
lines and based on the bulge luminosity or stellar dispersion of their
host galaxies \citep[e.g.][]{Mathur01, Grupe04, Mathur12}; for a given
black hole mass, narrow-line Seyfert 1s tend to reside in galaxies
with more luminous bulges, often pseudo-bulges, that are most likely
gas-rich. Finally, most recently, \citet{Yao15} estimated the black
hole mass based on the short timescale X-ray variability observed in a
{\it Suzaku} X-ray spectrum from 2009. Based on the normalised excess
variance for a 40~ks binning of the $2-4$~keV~light curve, they get a
value of (0.86$^{+0.29}_{-0.27}$)$\times$10$^7$~solar masses using the
\citet{Ponti12} relationship for the reverberation-mapped AGN
sample. This value is very similar to our result for the 20~ks binning
of the {\sl NuSTAR} light curve and, within the $2\sigma$ error range,
also consistent with our result for a 40~ks time binning.

\section{The spectral energy distribution} \label{sed}

\begin{figure*}
\centerline{
\includegraphics[clip=true, bb=10 300 570 692, scale=0.52]{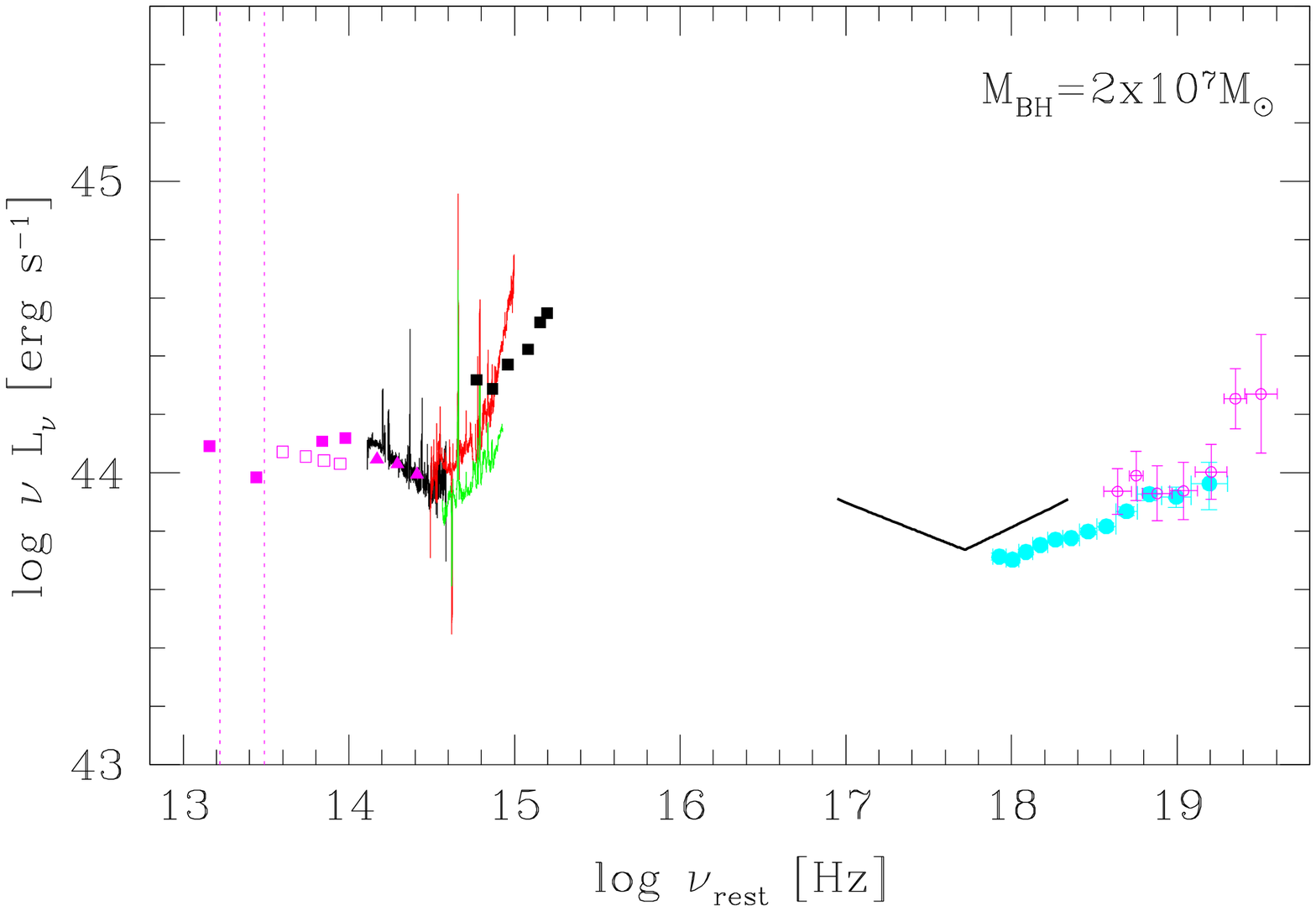}
\hspace*{-0.35cm}
\includegraphics[clip=true, bb=80 300 570 692, scale=0.52]{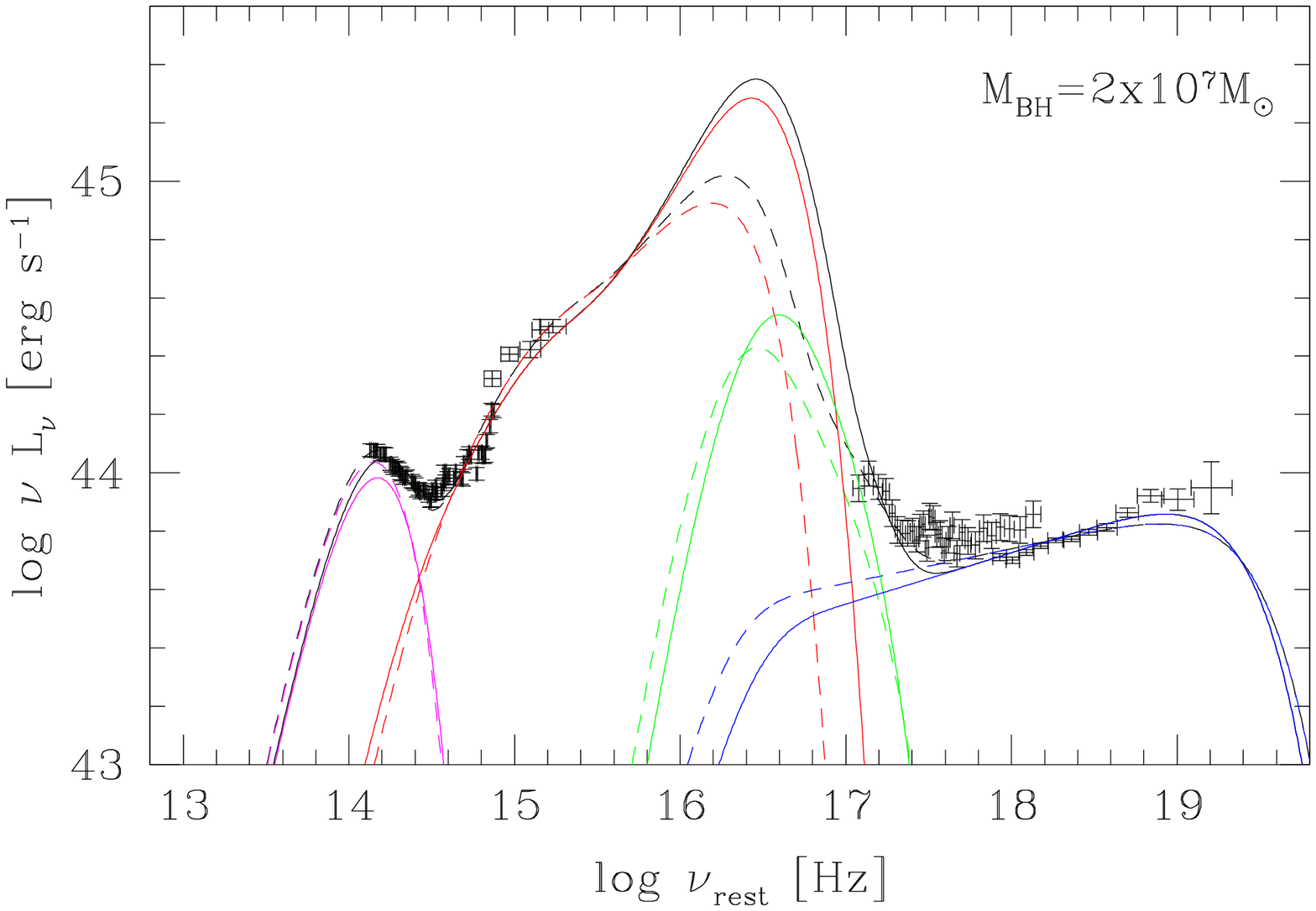}
}
\caption{\label{sedplot} Left-hand panel: Rest-frame spectral energy
  distribution for the source 1H~0323$+$342 based on {\it Gemini}
  GNIRS near-IR spectroscopy (black solid line), {\it Keck} optical
  spectroscopy (red solid line) and {\it NuSTAR} X-ray spectroscopy
  (cyan filled circles). The optical spectrum has been scaled up to
  match the flux of the near-IR spectrum in the overlap wavelength
  region. The optical spectrum from \citet{Marcha96} (green solid
  line) is included for comparison. The {\it Swift} optical/UV
  magnitudes close in time with the near-IR spectroscopy are shown as
  the black, filled squares and the best-fit to the co-added {\it
    Swift} X-ray spectrum as the thick, black solid line. The magenta
  filled triangles, filled and open squares and open circles indicate
  the 2MASS, WISE and {\it Spitzer} IRAC infrared photometry and the
  {\it Swift} BAT hard X-ray spectrum, respectively. The frequencies
  of the 10~$\mu$m and 18~$\mu$m dust silicate features are indicated
  by the vertical magenta dotted lines. Right-hand panel: Results from
  a fit of the accretion disc model of \citet{Done12, Done13} to the
  binned data (black points) for a non-rotating ($a^{\star}=0$) black
  hole (dashed lines) and a black hole with spin $a^{\star}=0.8$
  (solid lines) for a fixed mass of 2$\times$10$^7$~M$_\odot$. The
  total accretion disc spectrum including its reprocessed components
  is shown in black with the individual components as follows: thermal
  accretion disc spectrum (red), soft Comptonised component (green),
  and hard Comptonised component (blue). The additional blackbody for
  the hot dust component is shown as the magenta line.}
\end{figure*}

\subsection{The Eddington ratio} \label{Eddratio}

\begin{table*}
\caption{\label{modelpar} Best-fit parameter values for the accretion
  disc model of \citet{Done12, Done13} assuming a Schwarzschild
  ($a^{\star}=0$) and a Kerr ($a^{\star}=0.8$) black hole with a mass
  of 2$\times$10$^7$~M$_\odot$}
\begin{tabular}{lccccccccc}
\hline
& $L/L_{\rm Edd}$ & $\dot{M}$             & $\nu L_{\rm acc}$ & $r_{\rm cor}$ & $f_{\rm pl}$ & $\nu L_{\rm pl}$ & $T_{\rm dust}$ & $\nu L_{\rm dust}$ & $\chi^2_{\nu}$/dof \\
&                 & (M$_\odot$~yr$^{-1}$) & (erg~s$^{-1}$)    & ($r_{\rm g}$) &              & (erg~s$^{-1}$)   & (K)            & (erg~s$^{-1}$)     & \\
& (1)             & (2)                   & (3)               & (4)           & (5)          & (6)              & (7)            & (8)                & (9) \\
\hline
$a^{\star}=0$   & 0.55 & 0.41 & 1.86e$+$45 & 19 & 0.5 & 3.39e$+$44 & 1627 & 1.22e$+$44 & 4.92/98 \\
$a^{\star}=0.8$ & 1.00 & 0.37 & 3.39e$+$45 &  7 & 0.4 & 3.15e$+$44 & 1717 & 1.08e$+$44 & 4.51/98 \\
\hline
\end{tabular}

\parbox[]{14cm}{The columns are: (1) Eddington ratio; (2) accretion
  rate; (3) total accretion disc luminosity for the thermal component;
  (4) radius of the corona (in gravitational radii); (5) fraction of
  the power below the coronal radius that is reprocessed into the hard
  Comptonised component; (6) total luminosity of the hard Comptonised
  component; (7) blackbody temperature of the hot dust component; (8)
  total luminosity of the hot dust component; and (9) reduced $\chi^2$
  and number of degrees of freedom for the best-fit. We fixed the
  inclination of the accretion disc to an angle of $i=0^{\circ}$, the
  outer radius of the accretion disc to the self-gravity radius and
  the optical depth of the soft Comptonised component to a value of
  $\tau=15$. }

\end{table*}

With a reliable black hole mass estimate in hand, we are now in the
position to derive the Eddington ratio $L/L_{\rm Edd}$, where $L$ is
the total luminosity of the accretion disc and $L_{\rm Edd}$ is the
luminosity of an accretion disc accreting matter at the Eddington
limit for a given black hole mass. For this purpose, we have modelled
the observed SED of the source 1H~0323$+$342 from near-IR to X-ray
frequencies (see Fig. \ref{sedplot}) using the energy-conserving
accretion disc model of \citet{Done12}, which was later revised by
\citet{Done13}. In short, the model has three components: (i) a
relativistic, geometrically thin, optically thick accretion disc,
which emits thermal (blackbody) radiation with a spectrum that
includes a colour correction term to account for the fact that the
disc is not fully thermalized at all radii; (ii) a soft X-ray excess
component attributed to low-temperature, optically thick
Comptonisation of inner disc photons; and (iii) an X-ray power-law
attributed to high-temperature, optically thin Comptonisation. This
model is incorporated in the \xspec~analysis package under the name
\donemodel. However, it is important to also include an inclination
dependence of the accretion disc emission and relativistic corrections
such as gravitational redshift. These effects are incorporated in the
code \newdonemodel, which we have used here.

We have performed the accretion disc fits including the following
data; the binned near-IR and optical spectrum, whereby we first
subtracted the \FeII~emission from the optical spectrum and excluded
bins that contained strong emission lines and sampled the Balmer
continuum in the blue, the {\it Swift} UVOT magnitudes from the 2015
epoch with the exception of the $V$ magnitude, which falls on a strong
emission line, the co-added {\it Swift} XRT spectrum binned to a
minimum of 100 counts per energy bin and the {\it NuSTAR} spectrum. We
have assumed the two cases of a non-rotating Schwarzschild black hole,
i.e. a dimensionless spin parameter of $a^{\star}=0$, and a rotating
Kerr black hole with $a^{\star}=0.8$. Furthermore, we have included an
additional blackbody in our accretion disc fits in order to
simultaneously model the hot dust emission in the near-IR. The
resulting best-fit values for the relevant model parameters are listed
in Table \ref{modelpar} and the fits are shown in Fig. \ref{sedplot},
right-hand panel. We have fixed three of the free parameters, namely,
the accretion disc inclination to an angle of $i=0^{\circ}$ (face-on
view), the outer radius of the accretion disc to the self-gravity
radius and the optical depth of the soft Comptonised component to a
value of $\tau=15$. The assumed accretion disc inclination angle is
close to the value range of $i=4^{\circ}-13^{\circ}$ recently obtained
by \citet{Fuhr16} for the orientation of the radio jet based on the
apparent superluminal speeds of individual radio components they see
on high-spatial resolution Very Large Baseline Array (VLBA)
images. For the optical depth, we have assumed the mean value obtained
by \citet{Done12} for their modelling of the mean optical to X-ray AGN
SEDs of \citet{Jin12}.

The most important parameter that we would like to be able to
constrain from our SED fitting is spin. This would not only determine
the Eddington ratio of the source and so help establish if it is a
high accretor as usually found for the class of narrow-line Seyfert
1s, but it would be especially important in this case, since a high
black hole spin contrary to the zero spin usually found for
radio-quiet narrow-line Seyfert 1s \citep[e.g.][]{Done13} might
explain why the source 1H~0323$+$342 has a relativistic jet and so is
a strong $\gamma$-ray emitter \citep{Done16}. Although, it could be
that a high spin alone is not a sufficient condition for the
production of relativistic jets \citep{Fosch16b}. In the absence of
far-UV data, which is generally not available for AGN, the spin value
can in principle be constrained if a soft X-ray excess is
detected. However, the spectral slope of this component and its
frequency coverage are important, too, since only a very steep soft
X-ray excess well-sampled down to the lowest X-ray energies can
exclude spin values $a^{\star}>0$. In our case, we detect a soft X-ray
excess component in the co-added {\it Swift} spectrum, but its
frequency coverage reaches down to only $\sim 0.4$~keV, which is not
low enough to differentiate between zero spin and a spin value up to
$a^{\star}=0.8$. Therefore, we can constrain the Eddington ratio only
to a range of values, namely, $L/L_{\rm Edd} \sim 0.5 - 1$. However,
the model cannot fit the data with spin values of $a^{\star}>0.8$,
since these solutions strongly overpredict the observed soft X-ray
flux. Though we note that these solutions would also be of a
super-Eddington nature, in which case energy conservation may not be
appropriate due to losses via strong winds and advection \citep[see
  e.g.,][]{Jin16, Done16}. Thus, extremely high spins cannot be ruled
out if they are accompanied by the expected strong losses for highly
super-Eddington flows.

\subsection{The jet contribution} \label{jet}

We have interpreted the entire SED from near-IR to hard X-ray
frequencies as emission that is unassociated with the radio jet. In
particular, the synchrotron jet emission is expected to peak at
infrared frequencies and the associated inverse Compton emission,
which is assumed to dominate at $\gamma$-ray frequencies, might be
detecteable already at hard X-ray frequencies. In Fig. \ref{sedplot},
left panel, we plot in addition to our near-IR spectroscopy also the
photometry from the Two Micron All-Sky Survey \citep[2MASS;][]{2MASS}
Point Source Catalogue, the Wide-field Infrared Survey Explorer
\citep[WISE;][]{WISE} all-sky survey and the Spitzer Enhanced Imaging
Products (SEIP) source list. First, we note that the optical and
near-IR spectrum of the source 1H~0323$+342$ form together a butterfly
shape around an inflection point with a rest-frame wavelength of
1~$\mu$m, which is typical of radio-quiet AGN and generally
interpreted as the sum of emission from the (decreasing) thermal
accretion disc spectrum and the (rising) hot dust emission
\citep[e.g.,][]{Carl87, Glik06, L11a}. Secondly, no signifcant
variability is detected in the near-IR between our GNIRS spectroscopy
and the 2MASS photometry, which was obtained about 17 years earlier on
January 20, 1998. Neither is strong variability observed in the mid-IR
between the {\it Spitzer} IRAC photometry taken on September 18, 2008
and the WISE photometry taken on February 10/11, 2010. This points to
a dusty torus origin of the infrared emission rather than the
synchrotron emission from the relativistic jet, which was what was
modelled so far in this frequency range \citep[e.g.][]{Abdo09,
  Paliya14, Yao15}. Finally, the emission upturn evident between the
two WISE photometry points at the longest wavelengths is most likely
due to the fact that they are sampling the two dust silicate features
at rest-frame wavelengths of 10~$\mu$m and 18~$\mu$m (vertical magenta
dotted lines in Fig. \ref{sedplot}). A strong emission upturn is often
observed between these two features when in emission rather than in
absorption \citep[see, e.g., radio-loud quasars in][]{L10b}.

The inverse Compton emission from the jet most likely starts to
dominate over the accretion disc Comptonised power-law at the highest
energies sampled by \mbox{\it NuSTAR}. We find that the {\it NuSTAR}
data points at energies $\ga 22$~keV lie significantly above the
best-fit Comptonised power-law, which we also found evidenced at a low
significance level when fitting the data with a sum of two power-laws
(see Section \ref{nustar}). The prominance of the jet is then clearly
established at energies $\ga 90$~keV ($\nu \ga 10^{19.2}$~Hz) as shown
in Fig. \ref{sedplot}, left panel, where we have added the hard X-ray
spectrum from the {\it Swift} BAT 70-month catalogue
\citep{SwiftBATcat} fitted with a single power-law. Whereas the
averaged {\it Swift} BAT flux is consistent with the {\it NuSTAR} data
in the frequency range of overlap, excess emission is observed at the
high-energy end of the spectrum. A jet dominance of the hard X-rays,
and even of X-ray energies as low as a few keV when the source is in a
high state, was also reported by \citet{Fosch09} and \citet{Fosch12}
(see their Fig. 1, left panel), who analysed the {\it INTEGRAL} IBIS
and {\it Swift} XRT and BAT data available at the time. However, we
note that the $2-10$~keV spectral index of our source is much flatter
than that usually expected for the coronal emission of high-Eddington
sources \citep{Done12, Shem08}. Then, if most of the X-ray emission
were from the jet rather than the accretion disc corona, it is
suprising that the X-ray variability is so similar to the expectations
of coronal variability (see Section \ref{xrayvar}). Nonetheless, this
could simply indicate that there is a tight linkage between the corona
and the jet. We will explore the jet contribution in detail in a
subsequent paper (Kynoch et al., in preparation).

\section{Summary and conclusions}

We have presented here recent quasi-simultaneous optical and near-IR
spectroscopy of high quality (high S/N and moderate spectral
resolution) for the source 1H~0323$+$342, which is the lowest-redshift
member of the rare class of $\gamma$-ray detected narrow-line Seyfert
1s. We have supplemented these observations with a {\it NuSTAR} X-ray
spectrum taken about two years earlier and constrained the black hole
mass based on several optical and near-IR broad emission lines as well
as the short timescale X-ray variability. With a reliable black hole
mass estimate in hand, we have derived the Eddington ratio based on a
detailed spectral fitting of our multiwavelength SED with the
accretion disc model of Done et al. Our main results can be summarised
as follows.

\vspace*{0.2cm}

(i) Our three estimates of the black hole mass based on the ionising
continuum luminosity and the width of the hydrogen broad emission
lines H$\alpha$, H$\beta$ and Pa$\alpha$ give a very small range of
values of $\sim 1.5 - 2.2 \times 10^7$~solar masses, with an average
value of $\sim 2 \times 10^7$~solar masses. This amazing consistency
between the three estimates is suprising, given that the relationships
which they are based on have uncertainties of the order of $\sim
40-50\%$ at the $1\sigma$~level.

(ii) We obtain a very good agreement between the black hole mass
estimates from the broad emission lines and those from the short-term
X-ray variability, which lie in the range of $\sim 1.0 - 1.7 \times
10^7$~solar masses. In addition, we have considered alternative
methods to estimate the black hole mass, which are based on the
dispersion instead of the FWHM of the broad emission line and the line
and X-ray luminosity instead of the ionising continuum luminosity. We
find in general a good agreement with our previous estimates, except
when using the emission line luminosities. These give black hole
masses in the range of $\sim 0.6-1.2 \times 10^7$~solar masses, which
are a factor of $\sim 2$ smaller than our other estimates.

(iii) The main aim of our SED fitting is to constrain the spin value,
which in turn determines the Eddington ratio and so the bolometric
luminosity. In agreement with previous studies, we find that, in the
absence of far-UV data, the spin value can in principle be constrained
if a soft X-ray excess is detected. We detect this component in our
co-added {\it Swift} spectrum, but its frequency coverage does not
reach low enough to differentiate between zero spin and a spin value
up to $a^{\star}=0.8$. Therefore, we constrain the Eddington ratio
only to a range of values of $L/L_{\rm Edd} \sim 0.5 - 1$. However, we
can exclude spin values of $a^{\star}>0.8$ and so a super-Eddington
nature of the source, since these solutions strongly overpredict the
observed soft X-ray flux.

\section*{Acknowledgments}

MJW and DK would like to thank the Observatoire de Paris for its
hospitality and support during some months of this work in the
framework of the Laboratoire Europ\'een Associ\'e (LEA) ELGA (European
Laboratory for Gamma-ray Astronomy). We thank Maria March\~a for
kindly making their optical spectrum available to us in electronic
format. HL is supported by a European Union COFUND/Durham Junior
Research Fellowship (under EU grant agreement number 267209). MB
acknowledges support from the National Aeronautics and Space
Administration (NASA) Headquarters under the NASA Earth and Space
Science Fellowship Program grant no. NNX14AQ07H. DK acknowledges the
receipt of an STFC studentship. This work is partly based on
observations obtained at the {\it Gemini} Observatory, which is
operated by the Association of Universities for Research in Astronomy,
Inc., under a cooperative agreement with the NSF on behalf of the
Gemini partnership: the National Science Foundation (United States),
the National Research Council (Canada), CONICYT (Chile), the
Australian Research Council (Australia), Minist\'erio da Ci\^encia,
Tecnologia e Inova\c{c}\~ao (Brazil) and Ministerio de Ciencia,
Tecnolog\'ia e Innovaci\'on Productiva (Argentina). Some of the data
presented herein were obtained at the W.M. Keck Observatory, which is
operated as a scientific partnership among the California Institute of
Technology, the University of California and NASA. The Observatory was
made possible by the generous financial support of the W.M. Keck
Foundation. This work made use of data from the {\it NuSTAR} mission,
a project led by the California Institute of Technology, managed by
the Jet Propulsion Laboratory and funded by NASA.

\bibliography{/Users/herminelandt/references}

\begin{thebibliography}{}

\bibitem[\protect\citeauthoryear{{Abdo} et~al.}{{Abdo} et~al.}{2009}]{Abdo09}
{Abdo}, A.~A., et~al. 2009, \apjl, 707, L142

\bibitem[\protect\citeauthoryear{{Acero} et~al.}{{Acero} et~al.}{2015}]{Fermi3}
{Acero}, F., et~al. 2015, \apjs, 218, 23

\bibitem[\protect\citeauthoryear{{Ant{\'o}n}, {Browne}, \&
  {March{\~a}}}{{Ant{\'o}n} et~al.}{2008}]{Anton08}
{Ant{\'o}n}, S., {Browne}, I.~W.~A.,  \& {March{\~a}}, M.~J. 2008, \aap, 490,
  583

\bibitem[\protect\citeauthoryear{{Arnaud}}{{Arnaud}}{1996}]{xspec}
{Arnaud}, K.~A. 1996, in Astronomical Society of the Pacific Conference Series,
  Vol. 101, Astronomical Data Analysis Software and Systems V, ed. G.~H.
  {Jacoby} \& J.~{Barnes}, 17

\bibitem[\protect\citeauthoryear{{Barr} \& {Mushotzky}}{{Barr} \&
  {Mushotzky}}{1986}]{Barr86}
{Barr}, P.,  \& {Mushotzky}, R.~F. 1986, \nat, 320, 421

\bibitem[\protect\citeauthoryear{{Baumgartner} et~al.}{{Baumgartner}
  et~al.}{2013}]{SwiftBATcat}
{Baumgartner}, W.~H., {Tueller}, J., {Markwardt}, C.~B., {Skinner}, G.~K.,
  {Barthelmy}, S., {Mushotzky}, R.~F., {Evans}, P.~A.,  \& {Gehrels}, N. 2013,
  \apjs, 207, 19

\bibitem[\protect\citeauthoryear{{Bentz} et~al.}{{Bentz}
  et~al.}{2013}]{Bentz13}
{Bentz}, M.~C., et~al. 2013, \apj, 767, 149

\bibitem[\protect\citeauthoryear{Bentz et~al.}{Bentz et~al.}{2009}]{Bentz09}
Bentz, M.~C., Peterson, B.~M., Netzer, H., Pogge, R.~W.,  \& Vestergaard, M.
  2009, ApJ, 697, 160

\bibitem[\protect\citeauthoryear{{Berton} et~al.}{{Berton}
  et~al.}{2016}]{Berton16}
{Berton}, M., {Foschini}, L., {Ciroi}, S., {Cracco}, V., {La Mura}, G., {Di
  Mille}, F.,  \& {Rafanelli}, P. 2016, \aap, 591, A88

\bibitem[\protect\citeauthoryear{Boroson \& Green}{Boroson \&
  Green}{1992}]{Bor92}
Boroson, T.~A.,  \& Green, R.~F. 1992, ApJS, 80, 109

\bibitem[\protect\citeauthoryear{{Carleton} et~al.}{{Carleton}
  et~al.}{1987}]{Carl87}
{Carleton}, N.~P., {Elvis}, M., {Fabbiano}, G., {Willner}, S.~P., {Lawrence},
  A.,  \& {Ward}, M. 1987, \apj, 318, 595

\bibitem[\protect\citeauthoryear{{Cooke} \& {Rodgers}}{{Cooke} \&
  {Rodgers}}{2005}]{gnirssoft}
{Cooke}, A.,  \& {Rodgers}, B. 2005, in Astronomical Society of the Pacific
  Conference Series, Vol. 347, Astronomical Data Analysis Software and Systems
  XIV, ed. {P.~Shopbell, M.~Britton, \& R.~Ebert}, 514

\bibitem[\protect\citeauthoryear{Dickey \& Lockman}{Dickey \&
  Lockman}{1990}]{DL90}
Dickey, J.~M.,  \& Lockman, F.~J. 1990, ARA\&A, 28, 215

\bibitem[\protect\citeauthoryear{{Done} et~al.}{{Done} et~al.}{2012}]{Done12}
{Done}, C., {Davis}, S.~W., {Jin}, C., {Blaes}, O.,  \& {Ward}, M. 2012,
  \mnras, 420, 1848

\bibitem[\protect\citeauthoryear{{Done} \& {Jin}}{{Done} \&
  {Jin}}{2016}]{Done16}
{Done}, C.,  \& {Jin}, C. 2016, \mnras, 460, 1716

\bibitem[\protect\citeauthoryear{{Done} et~al.}{{Done} et~al.}{2013}]{Done13}
{Done}, C., {Jin}, C., {Middleton}, M.,  \& {Ward}, M. 2013, \mnras, 434, 1955

\bibitem[\protect\citeauthoryear{{Elias} et~al.}{{Elias} et~al.}{2006}]{gnirs}
{Elias}, J.~H., {Joyce}, R.~R., {Liang}, M., {Muller}, G.~P., {Hileman}, E.~A.,
   \& {George}, J.~R. 2006, in Society of Photo-Optical Instrumentation
  Engineers (SPIE) Conference Series, Vol. 6269, Society of Photo-Optical
  Instrumentation Engineers (SPIE) Conference Series

\bibitem[\protect\citeauthoryear{{Foschini}}{{Foschini}}{2012}]{Fosch12}
{Foschini}, L. 2012, in Proceedings of the Conference "Nuclei of Seyfert
  galaxies and QSOs - Central engine \& conditions of star formation"
  (Max-Planck-Insitut f{\"u}r Radioastronomie (MPIfR), Bonn, Germany, 6-8
  November, 2012), 10

\bibitem[\protect\citeauthoryear{{Foschini}}{{Foschini}}{2016}]{Fosch16b}
{Foschini}, L. 2016, in 28th Texas Symposium on Relativistic Astrophysics,
  arXiv:1205.3128

\bibitem[\protect\citeauthoryear{{Foschini} et~al.}{{Foschini}
  et~al.}{2016}]{Fosch16a}
{Foschini}, L., et~al. 2016, in 28th Texas Symposium on Relativistic
  Astrophysics, arXiv:1602.08227

\bibitem[\protect\citeauthoryear{{Foschini} et~al.}{{Foschini}
  et~al.}{2009}]{Fosch09}
{Foschini}, L., {Maraschi}, L., {Tavecchio}, F., {Ghisellini}, G., {Gliozzi},
  M.,  \& {Sambruna}, R.~M. 2009, Advances in Space Research, 43, 889

\bibitem[\protect\citeauthoryear{{Fuhrmann} et~al.}{{Fuhrmann}
  et~al.}{2016}]{Fuhr16}
{Fuhrmann}, L., et~al. 2016, Research in Astronomy and Astrophysics, accepted,
  arXiv:1608.03232

\bibitem[\protect\citeauthoryear{{Glikman}, {Helfand}, \& {White}}{{Glikman}
  et~al.}{2006}]{Glik06}
{Glikman}, E., {Helfand}, D.~J.,  \& {White}, R.~L. 2006, \apj, 640, 579

\bibitem[\protect\citeauthoryear{{Green}, {McHardy}, \& {Lehto}}{{Green}
  et~al.}{1993}]{Green93}
{Green}, A.~R., {McHardy}, I.~M.,  \& {Lehto}, H.~J. 1993, \mnras, 265, 664

\bibitem[\protect\citeauthoryear{{Greene} \& {Ho}}{{Greene} \&
  {Ho}}{2005}]{Greene05}
{Greene}, J.~E.,  \& {Ho}, L.~C. 2005, \apj, 630, 122

\bibitem[\protect\citeauthoryear{{Greene} et~al.}{{Greene}
  et~al.}{2010}]{Greene10}
{Greene}, J.~E., et~al. 2010, \apj, 723, 409

\bibitem[\protect\citeauthoryear{{Grupe} \& {Mathur}}{{Grupe} \&
  {Mathur}}{2004}]{Grupe04}
{Grupe}, D.,  \& {Mathur}, S. 2004, \apjl, 606, L41

\bibitem[\protect\citeauthoryear{{Harrison} et~al.}{{Harrison}
  et~al.}{2013}]{nustar}
{Harrison}, F.~A., et~al. 2013, \apj, 770, 103

\bibitem[\protect\citeauthoryear{{Jin}, {Done}, \& {Ward}}{{Jin}
  et~al.}{2016}]{Jin16}
{Jin}, C., {Done}, C.,  \& {Ward}, M. 2016, \mnras, 455, 691

\bibitem[\protect\citeauthoryear{{Jin} et~al.}{{Jin} et~al.}{2012}]{Jin12}
{Jin}, C., {Ward}, M., {Done}, C.,  \& {Gelbord}, J. 2012, \mnras, 420, 1825

\bibitem[\protect\citeauthoryear{{Kaspi} et~al.}{{Kaspi}
  et~al.}{2000}]{Kaspi00}
{Kaspi}, S., {Smith}, P.~S., {Netzer}, H., {Maoz}, D., {Jannuzi}, B.~T.,  \&
  {Giveon}, U. 2000, \apj, 533, 631

\bibitem[\protect\citeauthoryear{{Kim}, {Im}, \& {Kim}}{{Kim}
  et~al.}{2010}]{Kim10}
{Kim}, D., {Im}, M.,  \& {Kim}, M. 2010, \apj, 724, 386

\bibitem[\protect\citeauthoryear{Landt et~al.}{Landt et~al.}{2011a}]{L11b}
Landt, H., Bentz, M.~C., Peterson, B.~M., Elvis, M., Ward, M.~J., Korista,
  K.~T.,  \& Karovska, M. 2011a, MNRAS, 413, L106

\bibitem[\protect\citeauthoryear{Landt et~al.}{Landt et~al.}{2008}]{L08a}
Landt, H., Bentz, M.~C., Ward, M.~J., Elvis, M., Peterson, B.~M., Korista,
  K.~T.,  \& Karovska, M. 2008, ApJS, 174, 282

\bibitem[\protect\citeauthoryear{Landt, Buchanan, \& Barmby}{Landt
  et~al.}{2010}]{L10b}
Landt, H., Buchanan, C.~L.,  \& Barmby, P. 2010, MNRAS, 408, 1982

\bibitem[\protect\citeauthoryear{Landt et~al.}{Landt et~al.}{2011b}]{L11a}
Landt, H., Elvis, M., Ward, M.~J., Bentz, M.~C., Korista, K.~T.,  \& Karovska,
  M. 2011b, MNRAS, 414, 218

\bibitem[\protect\citeauthoryear{{Landt} et~al.}{{Landt} et~al.}{2014}]{L14}
{Landt}, H., {Ward}, M.~J., {Elvis}, M.,  \& {Karovska}, M. 2014, \mnras, 439,
  1051

\bibitem[\protect\citeauthoryear{Landt et~al.}{Landt et~al.}{2013}]{L13}
Landt, H., Ward, M.~J., Peterson, B.~M., Bentz, M.~C., Elvis, M., Korista,
  K.~T.,  \& Karovska, M. 2013, MNRAS, 432, 113

\bibitem[\protect\citeauthoryear{{Le{\'o}n Tavares} et~al.}{{Le{\'o}n Tavares}
  et~al.}{2014}]{Leon14}
{Le{\'o}n Tavares}, J., et~al. 2014, \apj, 795, 58

\bibitem[\protect\citeauthoryear{{Madsen} et~al.}{{Madsen}
  et~al.}{2015}]{Madsen15}
{Madsen}, K.~K., et~al. 2015, \apjs, 220, 8

\bibitem[\protect\citeauthoryear{{Magdziarz} \& {Zdziarski}}{{Magdziarz} \&
  {Zdziarski}}{1995}]{pexrav}
{Magdziarz}, P.,  \& {Zdziarski}, A.~A. 1995, \mnras, 273, 837

\bibitem[\protect\citeauthoryear{March\~a et~al.}{March\~a
  et~al.}{1996}]{Marcha96}
March\~a, M. J.~M., Browne, I. W.~A., Impey, C.~D.,  \& Smith, P.~S. 1996,
  MNRAS, 281, 425

\bibitem[\protect\citeauthoryear{{Massey} \& {Gronwall}}{{Massey} \&
  {Gronwall}}{1990}]{Massey90}
{Massey}, P.,  \& {Gronwall}, C. 1990, \apj, 358, 344

\bibitem[\protect\citeauthoryear{{Mathur} et~al.}{{Mathur}
  et~al.}{2012}]{Mathur12}
{Mathur}, S., {Fields}, D., {Peterson}, B.~M.,  \& {Grupe}, D. 2012, \apj, 754,
  146

\bibitem[\protect\citeauthoryear{{Mathur}, {Kuraszkiewicz}, \&
  {Czerny}}{{Mathur} et~al.}{2001}]{Mathur01}
{Mathur}, S., {Kuraszkiewicz}, J.,  \& {Czerny}, B. 2001, \na, 6, 321

\bibitem[\protect\citeauthoryear{{McHardy} et~al.}{{McHardy}
  et~al.}{2006}]{McHardy06}
{McHardy}, I.~M., {Koerding}, E., {Knigge}, C., {Uttley}, P.,  \& {Fender},
  R.~P. 2006, \nat, 444, 730

\bibitem[\protect\citeauthoryear{{Mej{\'{\i}}a-Restrepo}
  et~al.}{{Mej{\'{\i}}a-Restrepo} et~al.}{2016}]{Mejia16}
{Mej{\'{\i}}a-Restrepo}, J.~E., {Trakhtenbrot}, B., {Lira}, P., {Netzer}, H.,
  \& {Capellupo}, D.~M. 2016, MNRAS, 460, 187

\bibitem[\protect\citeauthoryear{{Nandra} et~al.}{{Nandra}
  et~al.}{1997}]{Nandra97}
{Nandra}, K., {George}, I.~M., {Mushotzky}, R.~F., {Turner}, T.~J.,  \&
  {Yaqoob}, T. 1997, \apj, 476, 70

\bibitem[\protect\citeauthoryear{{Oke} et~al.}{{Oke} et~al.}{1995}]{lris}
{Oke}, J.~B., et~al. 1995, \pasp, 107, 375

\bibitem[\protect\citeauthoryear{{Onken} et~al.}{{Onken}
  et~al.}{2004}]{Onken04}
{Onken}, C.~A., {Ferrarese}, L., {Merritt}, D., {Peterson}, B.~M., {Pogge},
  R.~W., {Vestergaard}, M.,  \& {Wandel}, A. 2004, \apj, 615, 645

\bibitem[\protect\citeauthoryear{{Paliya} et~al.}{{Paliya}
  et~al.}{2014}]{Paliya14}
{Paliya}, V.~S., {Sahayanathan}, S., {Parker}, M.~L., {Fabian}, A.~C.,
  {Stalin}, C.~S., {Anjum}, A.,  \& {Pandey}, S.~B. 2014, \apj, 789, 143

\bibitem[\protect\citeauthoryear{Peterson}{Peterson}{1993}]{Pet93}
Peterson, B.~M. 1993, PASP, 105, 247

\bibitem[\protect\citeauthoryear{{Polletta} et~al.}{{Polletta}
  et~al.}{2007}]{Pol07}
{Polletta}, M., et~al. 2007, \apj, 663, 81

\bibitem[\protect\citeauthoryear{{Ponti} et~al.}{{Ponti}
  et~al.}{2012}]{Ponti12}
{Ponti}, G., {Papadakis}, I., {Bianchi}, S., {Guainazzi}, M., {Matt}, G.,
  {Uttley}, P.,  \& {Bonilla}, N.~F. 2012, \aap, 542, A83

\bibitem[\protect\citeauthoryear{{Shemmer} et~al.}{{Shemmer}
  et~al.}{2008}]{Shem08}
{Shemmer}, O., {Brandt}, W.~N., {Netzer}, H., {Maiolino}, R.,  \& {Kaspi}, S.
  2008, \apj, 682, 81

\bibitem[\protect\citeauthoryear{Skrutskie et~al.}{Skrutskie
  et~al.}{2006}]{2MASS}
Skrutskie, M.~F., et~al. 2006, AJ, 131, 1163

\bibitem[\protect\citeauthoryear{{Vaughan} et~al.}{{Vaughan}
  et~al.}{2003}]{Vaughan03}
{Vaughan}, S., {Edelson}, R., {Warwick}, R.~S.,  \& {Uttley}, P. 2003, \mnras,
  345, 1271

\bibitem[\protect\citeauthoryear{V\'eron-Cetty, Joly, \& V\'eron}{V\'eron-Cetty
  et~al.}{2004}]{Veron04}
V\'eron-Cetty, M.-P., Joly, M.,  \& V\'eron, P. 2004, A\&A, 417, 515

\bibitem[\protect\citeauthoryear{Vestergaard \& Peterson}{Vestergaard \&
  Peterson}{2005}]{Ves05}
Vestergaard, M.,  \& Peterson, B.~M. 2005, ApJ, 625, 688

\bibitem[\protect\citeauthoryear{{Wandel}, {Peterson}, \& {Malkan}}{{Wandel}
  et~al.}{1999}]{Wan99}
{Wandel}, A., {Peterson}, B.~M.,  \& {Malkan}, M.~A. 1999, \apj, 526, 579

\bibitem[\protect\citeauthoryear{{Wright} et~al.}{{Wright} et~al.}{2010}]{WISE}
{Wright}, E.~L., et~al. 2010, \aj, 140, 1868

\bibitem[\protect\citeauthoryear{{Xiao} et~al.}{{Xiao} et~al.}{2011}]{Xiao11}
{Xiao}, T., {Barth}, A.~J., {Greene}, J.~E., {Ho}, L.~C., {Bentz}, M.~C.,
  {Ludwig}, R.~R.,  \& {Jiang}, Y. 2011, \apj, 739, 28

\bibitem[\protect\citeauthoryear{{Yao} et~al.}{{Yao} et~al.}{2015}]{Yao15}
{Yao}, S., {Yuan}, W., {Komossa}, S., {Grupe}, D., {Fuhrmann}, L.,  \& {Liu},
  B. 2015, \aj, 150, 23

\bibitem[\protect\citeauthoryear{{Zhou} et~al.}{{Zhou} et~al.}{2007}]{Zhou07}
{Zhou}, H., et~al. 2007, \apjl, 658, L13

\end{thebibliography}

\clearpage
\appendix
\section{{\it Gemini} near-IR spectrum, {\it Keck} optical spectrum and {\it NuSTAR} light curve}

\begin{figure*} 
\centerline{
\includegraphics[scale=0.7, clip=true, bb= 35 235 580 700]{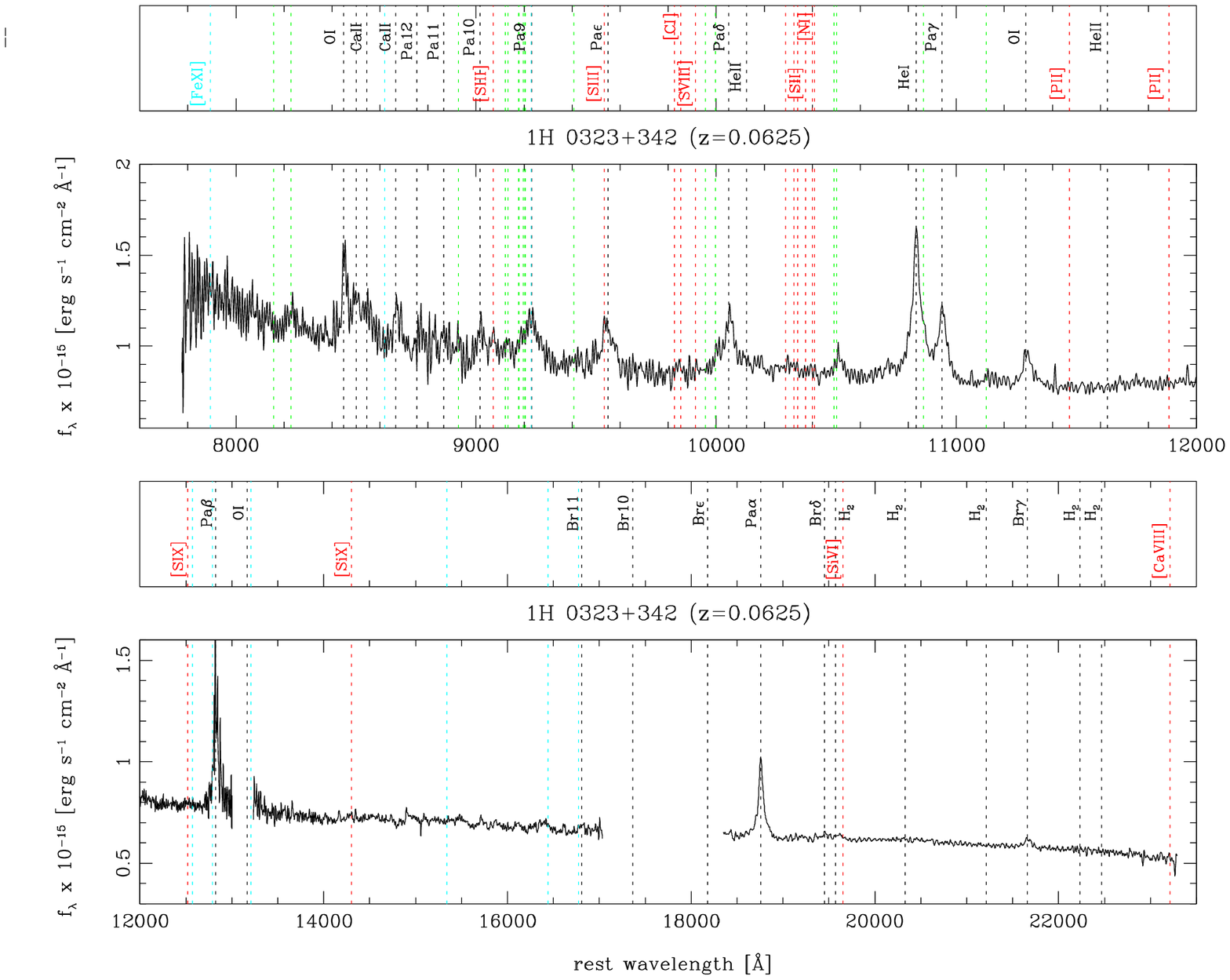}
}
\caption{\label{gnirsspec} Gemini GNIRS near-IR spectrum shown as
  observed flux versus rest-frame wavelength. Emission lines listed in
  Table 4 of \citet{L08a} are marked by dotted lines and labeled;
  black: permitted transitions, green: permitted \FeII~multiplets (not
  labeled), red: forbidden transitions and cyan: forbidden transitions
  of iron (those of \FeIIf~not labeled).}
\end{figure*}

\begin{figure*} 
\centerline{
\includegraphics[scale=0.5, clip=true, bb= 40 30 580 745, angle=-90]{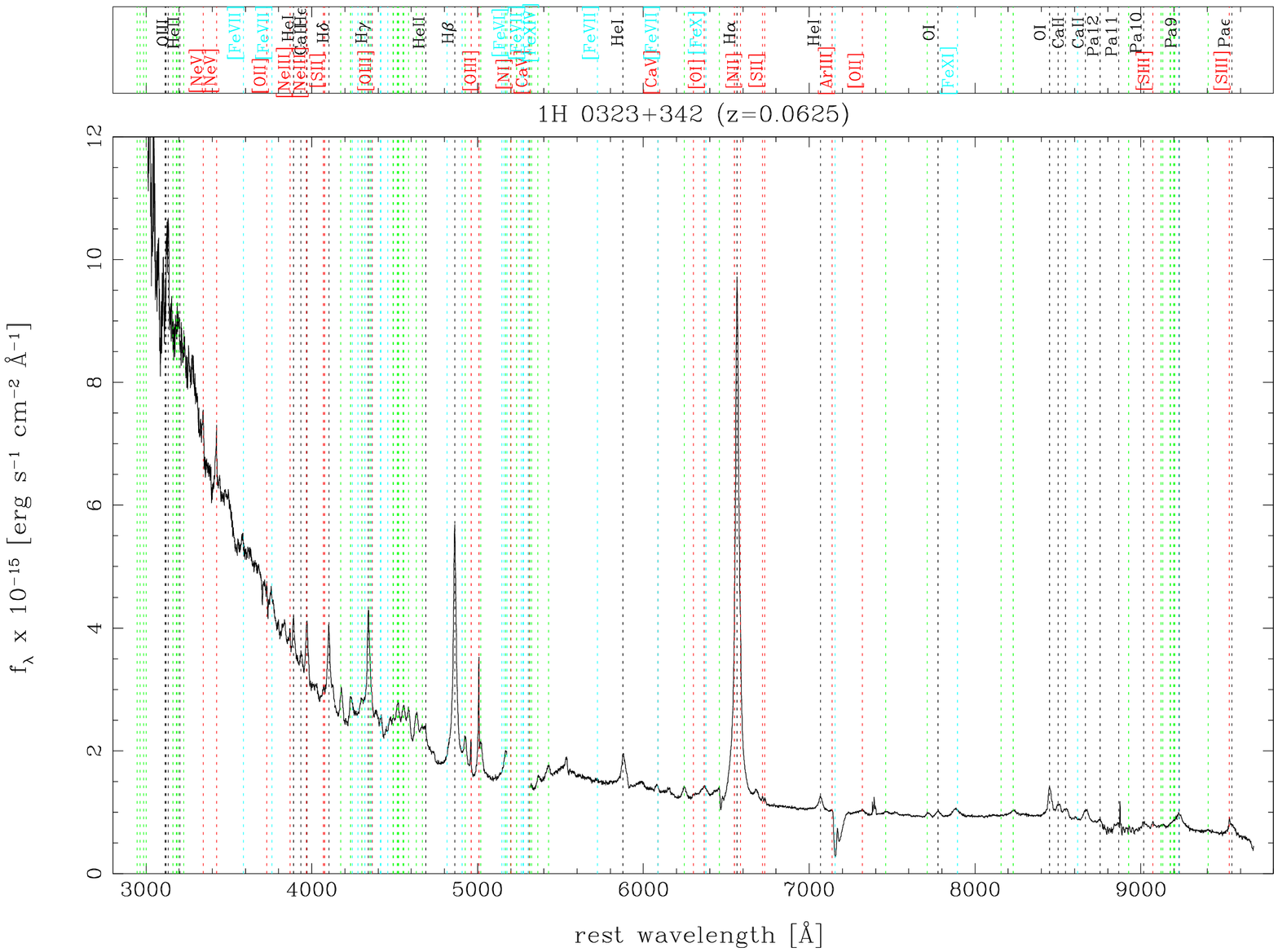}
}
\caption{\label{lrisspec} Keck LRIS optical spectrum shown as observed
  flux versus rest-frame wavelength. Emission lines labeled as in
  Fig. \ref{gnirsspec}.}
\end{figure*}

\begin{figure*}
\centerline{
\includegraphics[scale=0.5]{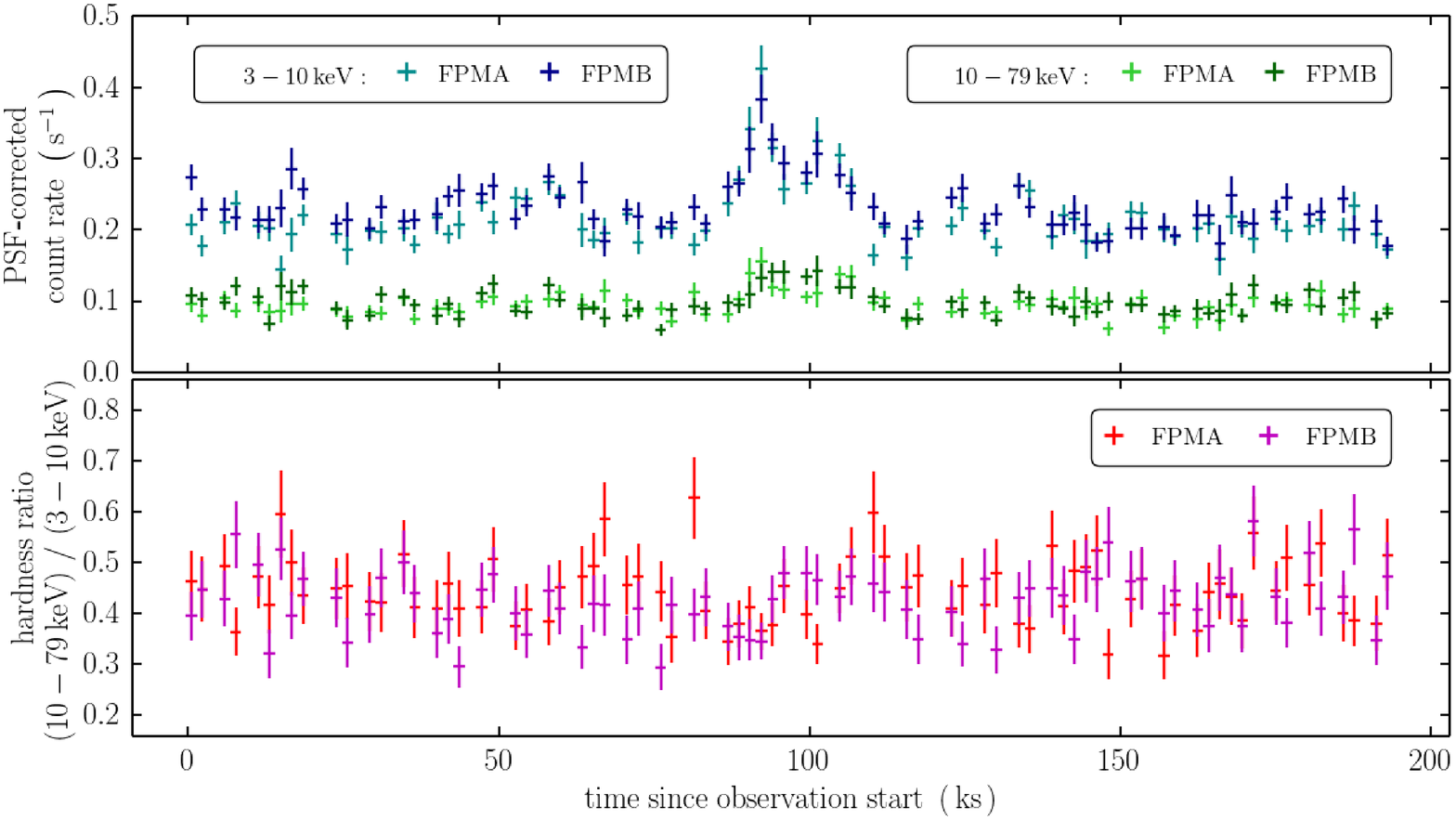}
}
\caption{\label{nustarlc} Top panel: {\it NuSTAR} light curve in bins
  of 30~minutes for the two focal plane modules A (FPMA) and B (FPMB)
  in the two energy bands 3--10~keV and 10--79~keV. Bottom panel: The
  ratio between the count rate in the 10--79~keV band and that in the
  3--10~keV band (i.e. the hardness ratio) for the two modules as a
  function of time. Note that the spectrum did not significantly
  change during the brief period of increased flux.}
\end{figure*}

\bsp
\label{lastpage}

\end{document}